\documentclass{river-journal}
\usepackage{rivps}
\usepackage{amsmath}
\usepackage{cite}
\usepackage{graphicx}
\usepackage{xcolor}
\usepackage{wrapfig}
\usepackage{psfrag}
\usepackage{afterpage}
\usepackage{subfigure}
\usepackage{comment}
\usepackage{pifont}
\usepackage{blindtext}
\usepackage{hyperref}

\par

\begin{document}
\begin{opening}
\title{JITA4DS: Disaggregated execution of Data Science Pipelines between the Edge and the Data Centre}
\author{Genoveva Vargas-Solar (1), Ali Akoglu (2),  Md Sahil Hassan (2)}
\institute{French Council of Scientific Research (CNRS)-LIRIS (1), \\
University of Arizona (2) \\ \texttt{genoveva.vargas-solar@liris.cnrs.fr, akoglu@arizona.edu,  sahilhassan@email.arizona.edu}}
\end{opening}

\runningtitle{JITA4DS}
\runningauthor{A. Akoglu, G. Vargas-Solar, M. S. Hassan}

\subsection*{Abstract}
This paper targets the execution of data science (DS) pipelines supported by data processing, transmission and sharing across several resources executing greedy processes. Current data science pipelines environments provide various infrastructure services with computing resources such as general-purpose processors (GPP), Graphics Processing Units (GPUs), Field Programmable Gate Arrays (FPGAs) and Tensor Processing Unit (TPU) coupled with platform and software services to design, run and maintain DS pipelines. These one-fits-all solutions impose the complete externalization of data pipeline tasks. However, some tasks can be executed in the edge, and the backend can provide just in time resources to ensure ad-hoc and elastic execution environments.

This paper introduces an innovative composable "Just in Time Architecture" for configuring DCs for Data Science Pipelines (JITA-4DS) and associated resource management techniques. JITA-4DS is a cross-layer management system that is aware of both the application characteristics and the underlying infrastructures to break the barriers between applications, middleware/operating system, and hardware layers. Vertical integration of these layers is needed for building a customizable Virtual Data Center (VDC) to meet the dynamically changing data science pipelines' requirements such as performance, availability, and energy consumption. Accordingly, the paper shows an experimental simulation devoted to run data science workloads and determine the best strategies for scheduling the allocation of resources implemented by JITA-4DS.



\section{Introduction}
Data infrastructures such as Google, Amazon, eBay, and E-Trade are powered by data centers (DCs) with tens to hundreds of thousands of computers and storage devices running complex software applications. 
Existing IT architectures are not designed to provide an agile infrastructure to keep up with the rapidly evolving next-generation mobile, big data, and data science pipelines demands. These applications are distinct from the "traditional" enterprise ones because of their size, dynamic behavior, and nonlinear scaling and relatively unpredictable growth as inputs being processed. Thus, they require continuous provisioning and re-provisioning of DC resources~\cite{chen19acm,kannan19eurosys,xu2015enreal} given their dynamic and unpredictable changes in the Service Level Objectives (SLOs) (e.g., availability response time, reliability, energy).  

This paper targets the execution of data science (DS) pipelines supported by data processing, transmission and sharing across several resources executing greedy processes. {A Data Science Pipeline consists of a set of data processing tasks organized as a data flow defining the data dependencies among the tasks and a control flow defining the order in which tasks are executed.}  Current data science pipelines environments promote high-performance cloud platforms as backend support for completely externalizing their execution. These platforms provide various infrastructure services with computing resources such as general-purpose processors (GPP), Graphics Processing Units (GPUs), Field Programmable Gate Arrays (FPGAs) and Tensor Processing Unit (TPU) coupled with platform and software services to design, run and maintain DS pipelines. These one-fits-all solutions impose the complete externalization of data pipeline tasks that assume (i) reliable and completely available network connection; (ii) can be energy and economically consuming, allocating large scale resources for executing pipelines tasks. However, some tasks can be executed in the edge, and the backend can provide just in time resources to ensure ad-hoc and elastic execution environments.

Our research investigates architectural support, system performance metrics, resource management algorithms, and modeling techniques to enable the design of composable (disaggregated) DCs. 
The goal is to design an innovative composable "Just in Time Architecture" for configuring DCs for Data Science Pipelines (JITA-4DS) and associated resource management techniques. DCs utilize a set of flexible building blocks that can be dynamically and automatically assembled and re-assembled to meet the dynamic changes in workload's Service Level Objectives (SLO) of current and future DC applications.
DCs configured using JITA-4DS provide ad-hoc environments efficiently and effectively meet the continuous changes in data-driven applications or workloads (e.g., data science pipelines). 
To assess disaggregated DC's, we study how to model and validate their performance in large-scale settings. 
Accordingly, the paper shows an experimental simulation devoted to run data science workloads and determine the best strategies for scheduling the allocation of resources implemented by JITA-4DS.


The remainder of the paper is organized as follows. Section \ref{sec:related} discusses related work identifying certain drawbacks and issues we believe remain open. Section \ref{sec:edgepipeline} JITA-4DS the just in time edge-based data science pipeline, execution environment proposed in this paper. Section \ref{sec:emulation} describes preliminary results regarding the use of JITA-4DS for executing data sciencce workloads. Finally Section \ref{sec:conclusion} concludes the paper and discusses future work.

\section{Related Work}\label{sec:related}

The work introduced in this paper is related to two types of approaches: (i) disaggregated data centers willing to propose alternatives to one fits all architectures; and (ii) data science pipelines' execution platforms relying on cloud services for running greedy data analytics tasks.

\subsection{Disaggregated data centers}
Disaggregation of IT resources has been proposed as an alternative configuration for data centers. Compared to the monolithic server approach, in a disaggregated data center, CPU, memory and storage are separate resource blades interconnected via a network. The critical enabler for the disaggregated data center is the network and management software to create the logical connection of the resources needed by an application \cite{7842314}.
The industry has started to introduce systems that support a limited disaggregation capability. For example, the Synergy system by Hewlett Packard Enterprise (HPE) ~\cite{synergy}, and the Unified Computing System (UCS)~\cite{Cisco} M series servers by Cisco are two commercial examples of composable infrastructures. 

HPE Synergy allows the CPU to be decoupled from the storage and memory, but components remain physically close together. Liqid~\cite{liquid} provides proprietary software that dynamically composes physical servers from the pools of bare metal resources such as compute pool, graphics pool, storage pool, and networking pool. DriveScale’s composable infrastructure~\cite{drivescale} collocates the cluster of diskless compute units and storage units within a rack.
The work introduced in \cite{7842314} proposes a disaggregated data
center network architecture, with a scheduling algorithm designed for disaggregated computing.

\subsubsection{Data Science Environments}

{\em Data analytics stacks}
environments provide the underlying infrastructure for managing data, implementing data processing workflows to transform them, and executing data analytics operations (statistics, data mining, knowledge discovery, computational science processes). 
For example, the Berkeley Data Analytics Stack (BDAS) from the AMPLAb project is a multi-layered architecture that provides tools for virtualizing resources, addressing storage, data processing and querying as underlying tools for big data-aware applications. AsterixDB from the Asterix project is a full-fledged big data stack designed as a scalable, open-source Big Data Management System (BDMS \cite{BDMS}). 
%
%

{\em Cloud based Data Science Environments}
Three families of environments 
provide tools to explore, engineer and analyse data collections. 
(i) data analytics stacks; 
They are notebook oriented environments  externalised on the cloud. A notebook is a JSON document, following a versioned schema, and containing an ordered list of input/output cells which can contain code, text using Markdown \cite{markdown}
mathematics, plots and rich media. They provide data labs  and  environments with  libraries for defining and executing notebooks.  Examples of existing data labs are Kaggle \cite{kaggle}
and CoLab \cite{colab}
from Google, and Azure Notebooks from Microsoft Azure \cite{msazure}. 
%
%
\subsubsection{Externalised Data Programming environments}
 can be installed locally or externalized in clouds. This environments are based on the notion of notebook that 
 Thereby, experiments can become completely and absolutely replicable. Data Labs with Runtime Environments allow data scientists to find and publish data sets, explore and build models in a web-based data-science environment, and work with other data scientists and machine learning engineers. Data labs
 offer a public data platform, a cloud-based workbench for data science, and Artificial Intelligence experiments.

{\em Platforms for custom modelling}
provide a suite of machine learning tools allowing developers with little experience to train high-quality models.  Tools are provided as services by commercial cloud providers that include storage, computing support and environments for training and enacting greedy artificial intelligence (AI) models. The leading vendors providing this kind of platforms are Amazon Sage Maker, Azure ML Services, Google ML Engine and IBM Watson ML Studio. 

 All except Azure ML Service provide built-in machine learning and artificial intelligence algorithms, prediction tools like linear regression, and operation for processing data structures such as tabular data representations. They also support the most commonly used machine learning and artificial intelligence libraries and frameworks for executing tasks that can be wrapped as pipelines. 

{\em Machine Learning and Artificial Intelligence Studios}
  give  an interactive, visual workspace to  build, test, and iterate on analytics models and develop experiments.  
An experiment consists of data sets that provide data to analytical modules connected to construct an analysis model.
 An experiment can be created from scratch or derived from an existing sample experiment as a template. 
 Data sets  and analysis modules can be drag-and-dropped onto an interactive canvas, connecting them together to form an experiment, which can be executed on  a machine learning runtime (cloud)  environment. 
 
{\em Machine learning runtime environments} provide the tools needed for executing machine learning workflows, including data stores, interpreters and runtime services like Spark, Tensorflow and Caffe for executing analytics operations and models.
The most prominent studios are, for example, Amazon Machine Learning, Microsoft Artificial Intelligence and Machine Learning Studio, Cloud Auto ML,  Data Bricks  ML Flow and IBM Watson ML Builder. 

 Each provide different families of built-in models such as classification, regression, clustering, anomaly detection, recommendation and ranking.
 
\subsubsection{Serverless and Edge based Approaches}
Serverless computing is a  computing execution model where the cloud provider allocates machine resources on demand. Serverless computing does not hold resources in volatile memory; computing is  done in short bursts with the results persisted to storage. Developers of serverless applications do not address capacity planning, configuration, management, maintenance, fault tolerance, or scaling of containers, VMs, or physical servers. Serverless vendors offer compute runtimes, also known as Function as a Service (FaaS) platforms, which execute application logic but do not store data. Kubeless and Fission are two Open Source FaaS platforms which run with Kubernetes.

Amazon AWS Lambda proposes an abstract serverless computing model supported by additional AWS serverless tools such as AWS Serverless Application Model (AWS SAM) Amazon CloudWatch. Google Cloud Platform created Google Cloud Functions. IBM offers IBM Cloud Functions in public, IBM Cloud and IBM Cloud Code Engine. Microsoft Azure offers Azure Functions, offered both in the Azure public cloud or on-premises via Azure Stack. Cloudflare offers Cloudflare Workers.

Edge computing \cite{davis2004edgecomputing} is a distributed computing paradigm that brings computation and data storage closer to the sources of data.
The increase of devices at the edge of the network produces a massive amount of data - storing and using all that data in cloud data centers pushes network bandwidth requirements to the limit.
The aim is to move the computation away from data centers towards the edge of the network, exploiting smart objects, mobile phones, or network gateways to perform tasks and provide services on behalf of the cloud \cite{merenda2020edge}. 
One definition of edge computing is any computer program that delivers low latency nearer to the requests \cite{garcia2015edge}.  Moving services to the edge enables content caching, service delivery, persistent data storage resulting in better response times and transfer rates.

\subsubsection{Discussion}
Machine learning studios address the analytics and data management divide with integrated backends for efficient execution of analytics activities pipelines allocating the necessary infrastructure  (CPU, FPGA, GPU, TPU) and platform (Spark, Tensorflow)  services.
These descriptions provide insight to data scientists about the size of the data collections, licences, provenance as well as data structure and content distributions that can be visually observed. 
Curated data collections associated with a search engine can be shared and used in target data science projects. Data labs offer storage space often provided by a cloud vendor (e.g., users of CoLab use their google drive storage space for data collections, notebooks and results). Execution environments associate computing resources for executing notebooks that use curated data collections.
Other open and private spaces can be coupled and used as persistence support for decoupling the enactment environment from the data persistence support (e.g., data can be initially stored in github and then uploaded to the data lab for analytics purposes). 
In both cases available space depends on the type of subscription to the data lab. 
Machine learning studios address the analytics and data management divide with integrated backends for dealing with efficient execution of analytics activities pipelines allocating the necessary infrastructure  (CPU, FPGA, GPU, TPU) and platform (Spark, Tensorflow)  services.

These environments provide resources (CPU, storage and main memory) for executing data science tasks. These tasks are repetitive, process different amounts of data and require storage and computing support. Data science projects have life cycle phases that imply in-house small scale execution environments, and they can evolve into deployment phases where they can touch the cloud and the edge resources. Therefore, they require underlying elastic architectures that can provide resources at different scales. Disaggregated data centers solutions seem promising for them. Our work addresses the challenges implied when coupling disaggregated solutions with data science projects.

\section{JITA-4DS: Just in time Edge Based Data Science Pipelines Execution}\label{sec:edgepipeline}

\begin{figure*}[h]
   \centering
   \includegraphics[width=\linewidth]{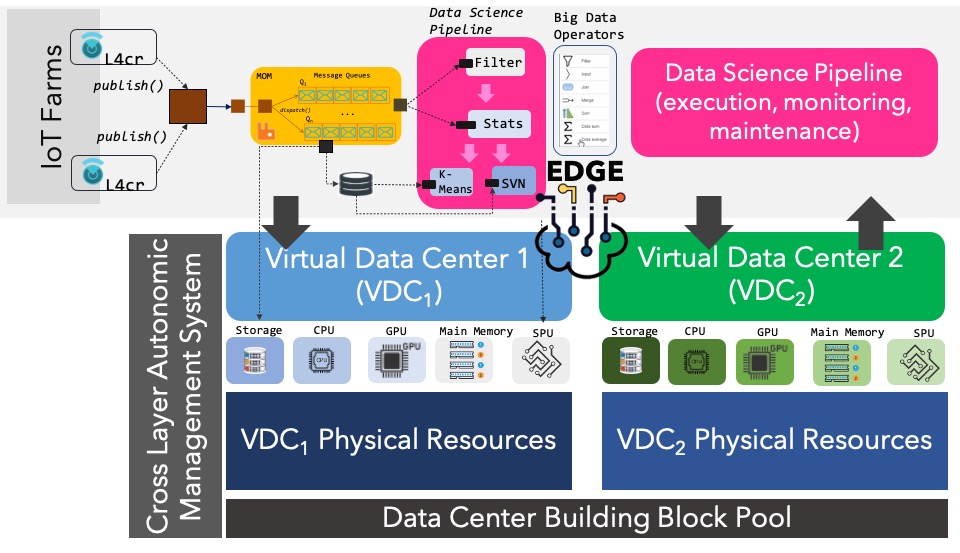}
    \caption{Just in Time Architecture for Data Science Pipelines - JITA-4DS}
   \label{fig:jita}
 \end{figure*}

The Just in Time Architecture for Data Science Pipelines (JITA-4DS), illustrated in Figure \ref{fig:jita}, is a cross-layer management system that is aware of both the application characteristics and the underlying infrastructures to break the barriers between applications, middleware/operating system, and hardware layers. Vertical integration of these layers is needed for building a customizable Virtual Data Center (VDC) to meet the dynamically changing data science pipelines' requirements such as performance, availability, and energy consumption. 

Our approach to design and development of the proposed JITA is illustrated in Figure \ref{fig:jita}. 

 The overall JITA design approach   focusses on the following interrelated research tasks:  
 \begin{enumerate}
     \item JITA middleware to support building VDCs that are optimized for each application type.
     \item  Development of AVDM system that will support efficient and dynamic management of VDC software and hardware building blocks. 
 \end{enumerate}

JITA-4DS benefits from virtualization techniques in the following aspects. 
Virtual Machine (VM) technologies such as VMware~\cite{vmware} and Xen~\cite{xen} provide a flexible management platform  useful for both the encapsulation of application execution environments and the aggregation and accounting of resources consumed by an application~\cite{machovec2017utility}. 
Consequently, VMs have been widely used to provide a layer that is well-positioned in the hardware/software stack of computer systems. 
Virtual Machine (VM) technologies provide fine-grain resource monitoring and control capabilities necessary for the JITA-4DS approach. Indeed, 

JITA-4DS fully exploits the virtualization from the virtual machine (VM) level into the VDC level  (e.g.,  fine-grain resource monitoring and control capabilities). JITA-4DS  can build a VDC that can meet the application SLO, such as execution performance and energy consumption to execute data science pipelines. The selected VDC, then, is mapped to a set of heterogeneous computing nodes such as GPPs, GPUs, TPUs, special-purpose units (SPUs) such as ASICs and FPGAs, along with memory and storage units. 

JITA-4DS encourages  a novel resource management methodology that is based on the time dependent value of service (VoS) metric to guide the assignment of resources to each VDC that maximizes the overall system-wide VoS metric. To predict the execution time and energy consumption of each application type, we use statistical and data mining techniques~\cite{kumbhare2017value,kumbhare2019adaptive,kumbhare2020dynamic,kumbhare2020value}, which represent the execution time and energy consumption as a function of the VDC resources. A complete study of these aspects for JITA-4DS have been described in \cite{akoglu2021putting}. 

DS pipelines running on top of JITA-4DS VDC's apply sets of big data processing operators to stored data and streams produced by the Internet of Things (IoT) farms (see the upper part of Figure \ref{fig:jita}). In the JITA-4DS approach, the tasks composing a 
data science pipeline are executed by services that implement big data operators. The objective is to execute as just in time edge-based processes (similar to lambda functions), and they interact with the VDC underlying services only when the process to execute needs more resources.  This means that services are running on the edge, on processing entities with different computing and storage capacities. They can totally or partially execute their tasks on the edge and/or on the VDC. 

This, in turn, creates the need for novel resource management approaches in streaming-based data science pipelines (see experimental discussion in Section \ref{sec:emulation}). These approaches should support and satisfy the data management strategy and stream exchange model between producers and consumers, invoke tasks with the underlying exchange model constraints on the compute and storage resources in the suitable form and modality and meet multi-objective competing performance goals.
Next, we describe the architecture of big data operators, and we show how they interact with the VDC. 
Later we will introduce our resource management approach for JITA-4DS.

The data management strategy stream exchange model. 
thanks to  the following interrelated research thrusts: (1) JITA Design Approach, and (2) Modeling, Analysis, and Simulation of JITA. 

\subsection
{Big Data/Stream producing and processing services}
\begin{figure}[t] 
\centering
\includegraphics[width=0.95\textwidth]{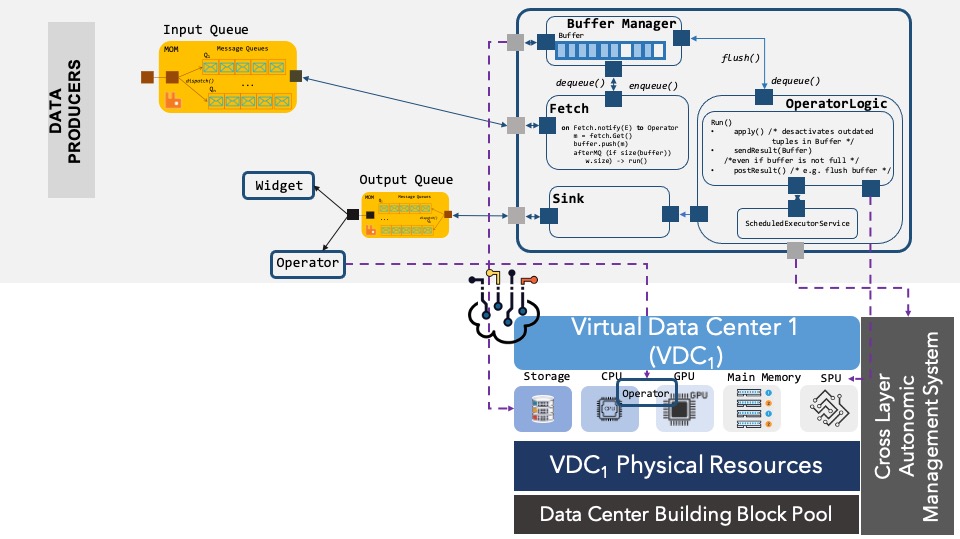}
\caption{Architecture of a  Big data/stream processing service}
\label{fig:micro-service}
\end{figure}

We assume that services that run on the edge produce and process data in batch or as streams. Data and stream processing services implement operators to support the analysis (machine learning, statistics, aggregation, AI) and visualise big data/streams produced in IoT environments.
As shown in Figure \ref{fig:jita}, data and stream producing services residing on edge rely on underlying message-based communication layers for transmitting them to processing and storage services. These services can reside on edge or a VDC.
A data/stream service implements simple or complex analytics big data operations (e.g., fetch, sliding window, average, etc.). Figure \ref{fig:micro-service} shows the general architecture of a streaming service.

A service consists of three key components, {\sf\small Buﬀer Manager, Fetch} and {\sf\small Sink}, and {\sf\small OperatorLogic}.
The service logic is based on a scheduler that ensures the recurrence rate in which the analytics operation implemented by the service is executed. Stream/data processing is based on unlimited consumption of data ensured by the component {\sf\small Fetch} that works if streams are notiﬁed by a producer. This speciﬁcation is contained in the logic of the components {\sf\small OperatorLogic} and {\sf\small Fetch}.
As shown in the ﬁgure, service communicates asynchronously with other micro-services using a message oriented middleware. 
As data is produced, the service fetches and copies the data to an internal buﬀer. Then, depending on its logic, it applies a processing algorithm and sends the data to the services connected to it. The general architecture of a service is specialized in concrete services implementing the most popular aggregation operations. 
These services can process data and streams on edge or a VDC. 

 These services are able to:
 \begin{itemize}
     \item [-] process data and streams on the edge or on a VDC on-line using tree window based strategies \cite{kramer2009semantics,golab2010data} (tumbling, sliding and landmark) well known in the stream processing systems domain;
 \item [-] combine stream histories with continuous ﬂows of streams of the same type (the average number of connections to Internet by Bob of the last month until the next hour).
 \end{itemize}

Since RAM assigned to a service might be limited, and in consequence its buﬀer, every service implements a data management strategy by collaborating with the communication middleware and with the VDC storage services to exploit buﬀer space, avoiding losing data, and processing and generating results on time.
 Big stream/data operators combine stream processing and storage techniques tuned depending on the number of things producing streams, the pace at which they produce them, and the physical computing resources available to process them online (on edge and VDC) and deliver them to the consumers (other services).
  Stores are distributively installed on edge and on the VDC.

Services adopt the tuple oriented data model as stream exchange model  with the IoT environment producing streams and the services. A stream is represented as a series of attribute value couples where values are of atomic types (integer, string, char, ﬂoat.
From a service point of view, a stream is a series of attribute value couples where values are of atomic types (integer, string, char, ﬂoat). 
We tuples are time-stamped, where  
assume that one of the attributes of the tuple corresponds to its time-stamp. 
the time-stamp represents the time of arrival of the stream to the communication infrastructure.
Data processing tasks navigate through the structure of the tuple for accessing attribute values. 

\subsection{Interval oriented storage support for consuming streams}
 Big stream/data operators combine stream processing and storage techniques tuned depending on the number of things producing streams, the pace at which they produce them, and the physical computing resources available for processing them on-line (on edge and VDC) and delivering them to consumers (other services).
  A service that aggregates historical data and streams includes a component named {\small\sf HistoricFetch}. 
  The component {\small\sf HistoricFetch} is responsible for performing a one-shot query for  retrieving stored data according to an input query. 
  Stores are distributively installed on edge and on the VDC.
 As described above, we have implemented a general/abstract micro-service that contains a {\small\sf Fetch} and {\small\sf Sink} micro-services. 
The historical fetch component has been specialized to interact with two stores that can be distributedly installed on the edge and on the VDC.
 \begin{itemize}
 \item [-]
 InfluxDB: a time series system accepting temporal queries, useful for computing time tagged tuples.
\item [-]	
 Cassandra: a key-value store that provides non-temporal read/write operations that might be interesting for storing huge quantities of data. 
 \end{itemize}
\subsection
{Edge based Data Science (DS) Pipelines}
are expressed by a series of data processing operations applied to streams/data stemming from things, stores or services. A DS pipeline is implemented by mashing up services implementing operators based on a composition operation that connects them by expressing a data flow (IN/OUT data). Aggregation (min, max, mean) and analytics (k-means, linear regression, CNN) services can be composed with temporal windowing services (landmark, sliding) that receive input data from storage support or a continuous data producer for instance, a thing.  The connectors are {\small\sf Fetch}, and {\small\sf Sink} services that determine the way services exchange data from/to things, storage systems, or other services (on-demand or continuous). Services can be hybrid (edge and VDC) services depending on the number of underlying services (computing, memory, storage) required.
To illustrate the use of a JITA-4DS, we introduce next a use case that makes full use of edge and VDC services configured ad-hoc for the analysis requirements.

 \subsection{Use Case: Analysing the connectivity of a connected society}
The experiment scenario aims at analyzing the connectivity of the connected society. 
The data is used then to answer queries such as:
 \begin{itemize}
 \item [-]
Connection speed of one internet provider:
{\em Am I receiving the network speed I am paying for all the time?}
\item [-] Internet availability with different Internet providers: 
{\em Which are the periods of the day in which I can upload/download files at the highest speed using different network providers?}
\end{itemize}

\begin{figure}[ht!] \centering
\includegraphics[width=0.85\textwidth]{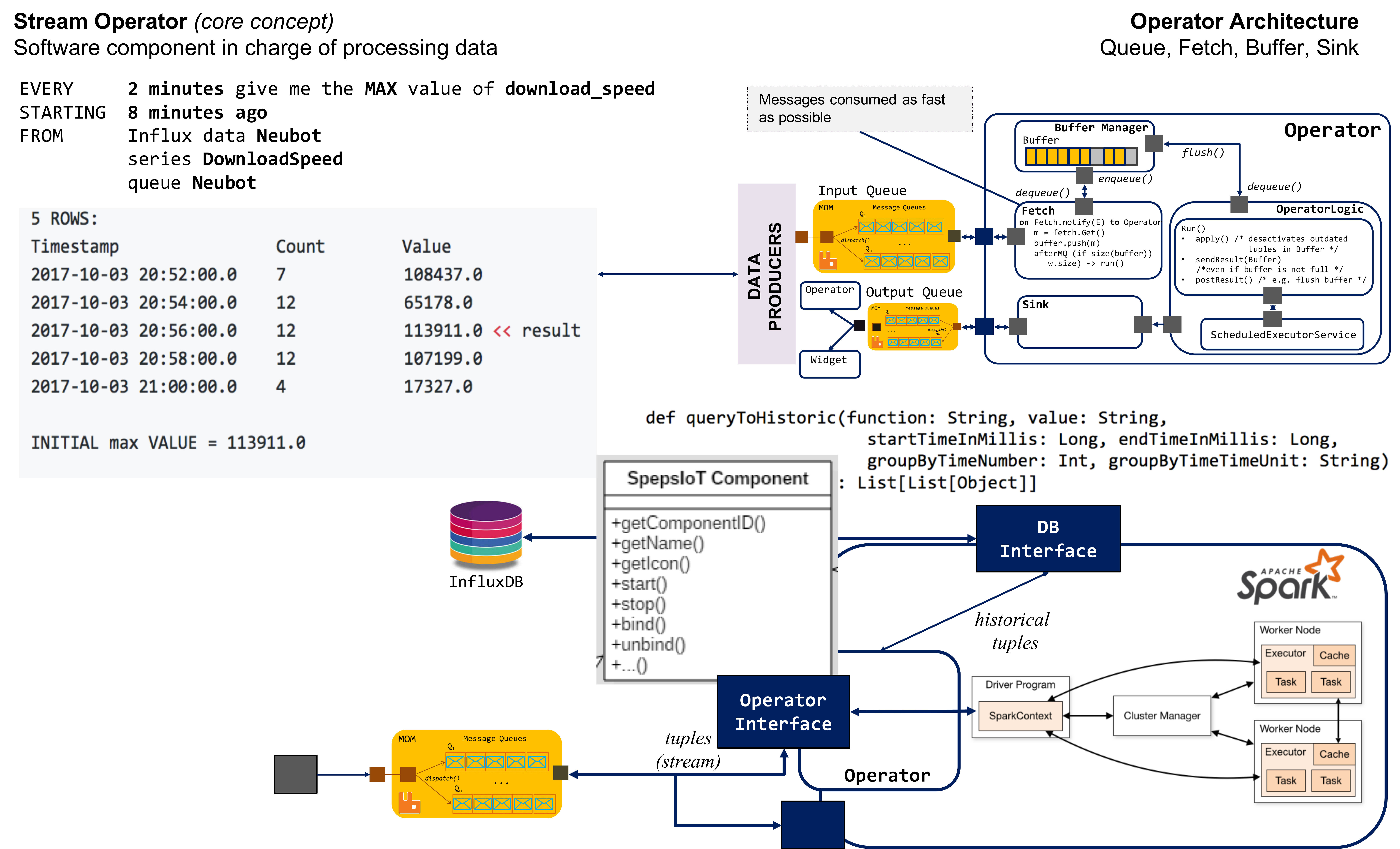}
\caption{Services architecture for analyzing society connectivity with the neubot dataset.}\label{fig:neubot-architecture}
\end{figure}
 
The data set used was produced in the context of the  Neubot project. Neubot is a  project on measuring the Internet from the edges by the Nexa Center for Internet and Society at Politecnico di Torino (https://www.neubot.org/).  It consists of network tests (e.g., download/upload speed over HTTP) realized by different users in different locations using an application that measures the network service quality delivered by different Internet connection types. The Neubot data collection was previously used in the context of the FP7 project S2EUNET.
The type of queries implemented as data science pipelines  were the following:
\begin{small}
\begin{verbatim}
EVERY 60 seconds compute the max value of download_speed 
of the last 3 minutes 
FROM 	cassandra database neubot series speedtests and streaming 
RabbitMQ queue neubotspeed

EVERY 	5 minutes compute the mean of the download_speed 
of the last 120 days 
FROM 	cassandra database neubot series speedtests and streaming 
rabbitmq queue neubotspeed

EVERY 30 seconds compute the mean value of upload_speed 
 starting 10 days ago 
FROM 	cassandra database neubot series speedtests and streaming
rabbitmq queue neubotspeed

\end{verbatim}
\end{small}

We built an IoT farm for deploying our experiment and implemented a distributed version of the IoT environment on a clustered version of RabbitMQ. This setting enabled us to address a scaleup setting regarding several data producers (things) deployed on edge. We installed aggregation operators as services distributed on the things and an edged execution environment deployed on servers deployed in different devices. The challenge is to consume streams, create a history of connectivity information, and then combine these voluminous histories with new streams to answer the queries. 
Depending on the observation window size, the services access the observations stored as post-mortem data sets from stores at the VDC level and connect to online producers currently observing their connections (on edge). %
For example, the second query observes a window of 10 days size. 
Our services could deal with histories produced in windows of size 10 days or even 120 days. Such massive histories could be combined with recent streams and produce reasonable response times (order of seconds).

\section{Runtime Emulation Environment}
\label{sec:emulation}

To conduct experiments on the DS workloads, we configure a runtime environment that  models the flexible heterogeneous resource pool and schedulers. This environment allows us sweep across the design search space of varying resource combinations and scheduling policies. For this purpose, we leverage the Compiler-Integrated Runtime  environment~\cite{mack2020user} that is an open-source Linux based user-space runtime framework. 

The integrated compiler in this environment converts the applications into a Directed Acyclic Graph (DAG) representation, where 
\begin{itemize}
    \item [-] A node is called a {\em task} representing a function used in the application domain (e.g., k-means).
    \item [-] An edge of the DAG represents a predecessor-successor dependency between the task nodes of an application.
\end{itemize}
 The compiler then generates a “flexible” binary structure for the runtime to invoke each DAG node on any of the available compute resources.

The runtime executes as a daemon process and consists of three key components:
\begin{itemize}
    \item [-] Application manager: parses the DAG and prepares handles for each kernel in the “flexible-binary” structure.
    \item [-] Workload manager: schedules the tasks of  applications on available PEs based on the user-defined scheduling policy and manages the data transfers to and from the PEs.
    \item [-] Resource manager: monitors the state of the PEs in the target hardware configuration and maintains coordination with the workload manager. 
\end{itemize}
Using this runtime environment, we modelled a hierarchical resource pool, described next.

\subsection{Hierarchical Resource Pool: a JITA4-DS instance}
\begin{figure}
    \centering
    \includegraphics[width=\linewidth]{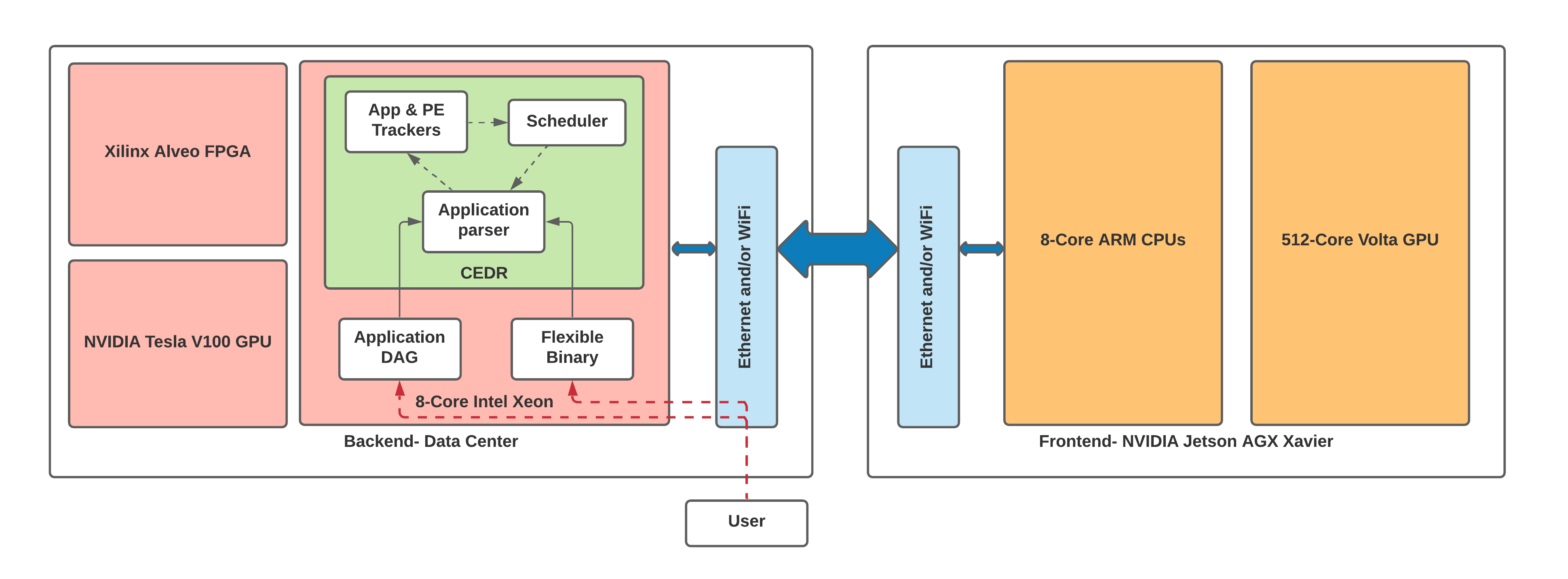}
    \caption{Diagram of Hierarchical architecture model using our runtime environment.}
    \label{fig:emulation_framework}
\end{figure}

  As shown in Figure \ref{fig:emulation_framework}, the hierarchical resource pool consists of two layers- the frontend, and the backend, with heterogeneous computing resources.
 
 The frontend represents the low-power computing resources present at the edge, such as Nvidia Volta GPU and ARM CPU cores. 
 
 The backend represents the high-performance compute resources such as Intel Xeon CPU cores, Nvidia Tesla V100 GPUs and Xilinx Alveo FPGA platforms. 
 
 The runtime is launched as a daemon process on one of the Xeon CPU cores on the backend platform and provided with the list of available frontend and backend resources. We assume that the user is submitting a job to the server (backend). The data flow of this job starts on the edge (frontend), where real-time data from the sensors are captured. This raw data are processed through the user-defined DS workflow. 

Job submission involves providing the application task flow DAG generated from the DS workflow and the flexible binaries to enable execution on the resources of the two-layer architecture. This DAG and flexible binary file are obtained by compiling the source code of DS workflow through the runtime environment-integrated compiler toolchain. At this stage, the runtime manager determines the mapping of individual tasks on the DAG to processing elements based on the state of computing resources at both the frontend (edge) and backend (server). 

This application submission process allows users to submit any desired number of application instances to the runtime, either all instances submitted at once or submitted with a periodic delay between each instance.

Upon receiving a job consisting of several tasks at the runtime, and after raw data from the edge sensors are captured, the runtime can schedule each task on either one of the heterogeneous resources available at the backend platform. This task involves requesting input data from and sending output data to the edge device or decide to schedule the task on one of the edge resources, which alleviates the overhead of communicating with the backend and exchanging necessary input-output data. 

To enable the runtime schedulers in making an informed decision, each task in the DAG file is assigned an expected execution time on the supported compute platforms based on historical data. The expected execution time of the tasks on the frontend resources denotes only the execution time, whereas, for backend resources, this time consists of both execution time and data communication overhead.

\subsection{Executing Data Science Workloads}
\begin{figure}[ht!] \centering
\includegraphics[width=0.85\textwidth]{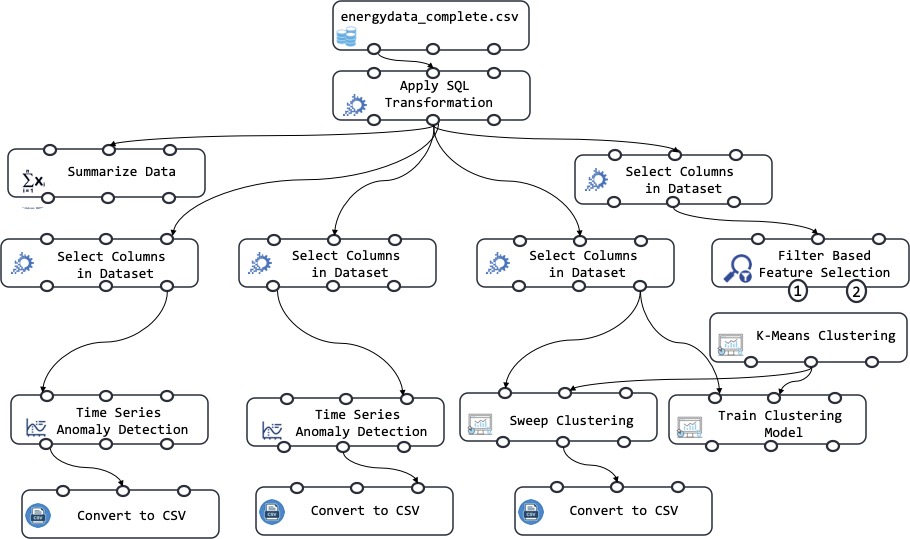}
\caption{Data Science Workload.}\label{fig:ds-workflow}
\end{figure}

We use the Data Science (DS) workloads as the application to run on the hierarchical architecture for our experiments. The DS workload (see Figure \ref{fig:ds-workflow}) consists of 16 task nodes, including: frequently used data science functions such as SQL Transform, data summarization, column selection in dataset, filter-based feature selection, k-means clustering, time series anomaly detection, sweep clustering, train clustering model etc. 

We assume that there is historical execution time data for each task node on each of the compute resources modeled. To further consider the communication cost, we assume the data rate of the communication channel to be  12 Mbps~\cite{4GSpeeds} and calculate the data transfer overhead by using the volume of input and output data of each task that needs to be transferred.

With the experimental setup, we can conduct elaborate experiments to get answers to some crucial questions. For example,  
\begin{itemize}
    \item [RQ$_1$] Is it beneficial to hand off all the tasks of given workloads to the backend resources.
    
    \item [RQ$_2$] Should some of the tasks be assigned to the frontend resources instead?
    
    \item [RQ$_3$] Can the communication cost of executing some tasks on the backend resources grow large enough for the runtime to decide to run it on a frontend resource?
    
\end{itemize}

In order to find these answers, we conduct two experiments on the hierarchical architecture model in our runtime environment. 

\subsubsection{Experiment 1: Sweeping across resource pool configurations}

In the first experiment, we fix the scheduling policy to Earliest Finish Time (EFT) and sweep across the combinatorial search space of resource pool configurations. Each resource pool configuration refers to a unique combination of front and backend heterogeneous resources with a specific count of each resource type. 

For our experiment, we vary the number of ARM cores and Xeon CPUs between 1 to 3 each, whereas we keep the number of Volta GPUs, Tesla V100 GPUs and Xilinx Alveo FPGAs fixed at 1. On each configuration, we run 100 instances of the DS workload submitted to the runtime at once. Upon receiving the execution time of the DS workload on all the tested configurations, we identify the configuration that gives the minimum execution time. Once this configuration is identified, we use it to conduct a second experiment. 

\begin{figure}
    \centering
    \includegraphics[width=\linewidth]{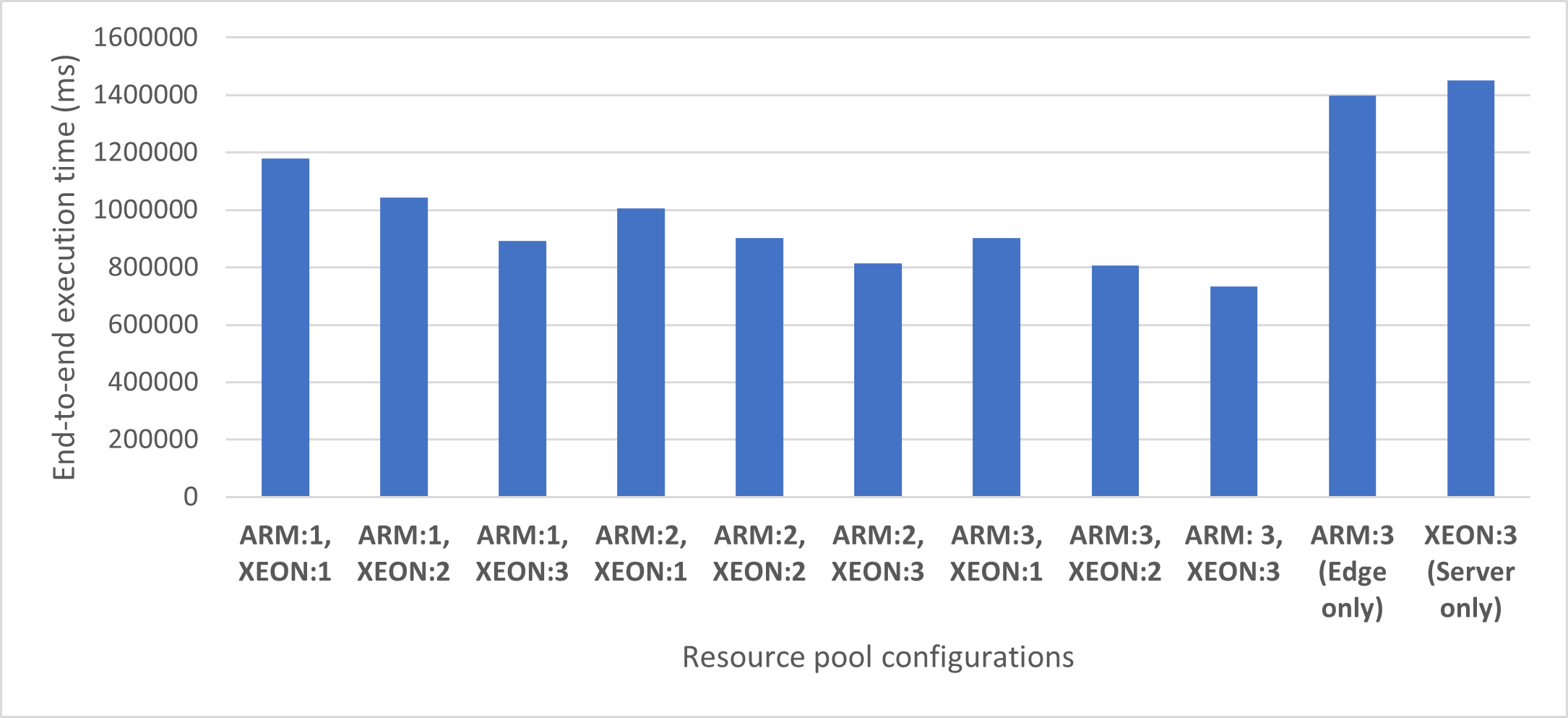}
    \caption{Execution time of 100 DS workflow instances for different resource pool configurations. First 9 columns of barchart also consist of 1 Nvidia Volta GPU (frontend), 1 Nvidia Tesla V100 GPU (backend) and 1 Xilinx Alveo FPGA (backend). The 10th configuration uses 3 ARM CPUs and 1 Volta GPU. The 11th configuration uses 3 Xeon CPUs, 1 Nvidia Tesla V100 GPU, and 1 Xilinx Alveo FPGA.}
    \label{fig:resource_pool_sweep}
\end{figure}

The result of experiment 1 is presented in Figure~\ref{fig:resource_pool_sweep}. This bar chart plots varying resource pool configurations along X-axis and their corresponding execution times for processing 100 instances of DS-workflow along Y-axis. The first 9 configurations (from left) include 1 Nvidia Volta GPU (at frontend), 1 Nvidia Tesla GPU and 1 Xilinx Alveo FPGA (at backend), along with the varying number of ARM and Xeon cores denoted by the label of the barplot. In the remaining two configurations, the \emph{Edge only} configuration consists of 3 ARM CPU cores and 1 Nvidia Volta GPU only, and executes all of the DS workload instances at the frontend. The \emph{Server only} configuration consists of 3 Xeon CPU cores, 1 Nvidia Tesla GPU and 1 Xilinx Alveo FPGA, and executes the entire application at the backend after collecting input data from frontend. In this plot, the two largest execution times are resulted by the \emph{Edge only} or \emph{Server only} configurations, whereas the remaining configurations consisting of a mixture of edge and server resources demonstrate lower execution times. The \emph{Edge only} configuration consists of low-power compute resources, which cause the overall workload execution on the edge slower. On the other hand, the \emph{Server only} configuration relies on the frontend to send larger amount of input data at the very beginning of workload execution, which increases the execution time significantly. Another observation we make here is that, with increasing number of available parallel resources (CPU, GPU or FPGA), the execution time reduces. Therefore, the lowest execution time is delivered by the configuration which holds the maximum number of resources of all types considered in this experiment (3 ARM, 1 Volta, 3 Xeon, 1 Tesla, 1 Alveo). Hence we choose this configuration to be used in conducting the second experiment.

Depending on the above observations, we can answer the research questions as follows-

\begin{itemize}
    \item [RQ$_1$] In terms of execution time, it is not beneficial to hand off all the tasks of a given workload to the backend resources. The initial data transfer overhead of large raw data from frontend to backend increases the execution time significantly, compared to if the data was to be used to run all the tasks at the frontend. Additionally, it is not beneficial to execute the entire workload on the frontend, as the frontend is constrained in terms of computation power. Hence, a mixture of both frontend and backend resources can be more beneficial.
    
    \item [RQ$_2$] Yes, to reduce the execution time, the data transfer overhead should be reduced, therefore some of the tasks should be executed at the frontend. 
    
    \item [RQ$_3$] Yes, in some scenarios the communication cost of transferring the data from frontend to backend might grow large enough for the runtime to decide to run those tasks on the frontend. This can be seen in the configurations with a mixture of frontend and backend resources. In these configurations the runtime decides to offload some tasks onto the frontend resources, which eventually reduces the execution time of the workload significantly (by upto 57\%) compared to running the entire workload on backend.
    
\end{itemize}

\subsubsection{Experiment 2: Sweeping across scheduling policies}

In this experiment, we sweep across different scheduling policies by running 100 instances of the DS workload submitted at once, using three different scheduling policies- ‘EFT’, ‘Earliest Task First (ETF)’, and ’Round Robin (RR)'. The EFT and ETF are sophisticated scheduling policies that take into account the hierarchy of the resource pool, expected execution time and data communication overhead of the underlying tasks in order to make scheduling decisions, while the RR is a simple scheduler that assigns tasks to resources in a round robin manner. We aim to identify the scheduling policy that gives the lowest execution time and highest mean resource utilization on the optimal configuration from experiment 1. 

This configuration includes 3 ARM CPU cores and 1 Nvidia Volta GPU at the frontend, along with 3 Xeon CPU cores, 1 Nvidia Tesla GPU and 1 Xilinx Alveo FPGA at the backend. We identify the resource utilization as the fraction of application execution time during which a particular resource is busy executing tasks. We take the mean of resource utilization of all resources to obtain the mean resource utilization.

\begin{figure}
    \centering
    \subfigure[]{\includegraphics[width=0.45\textwidth]{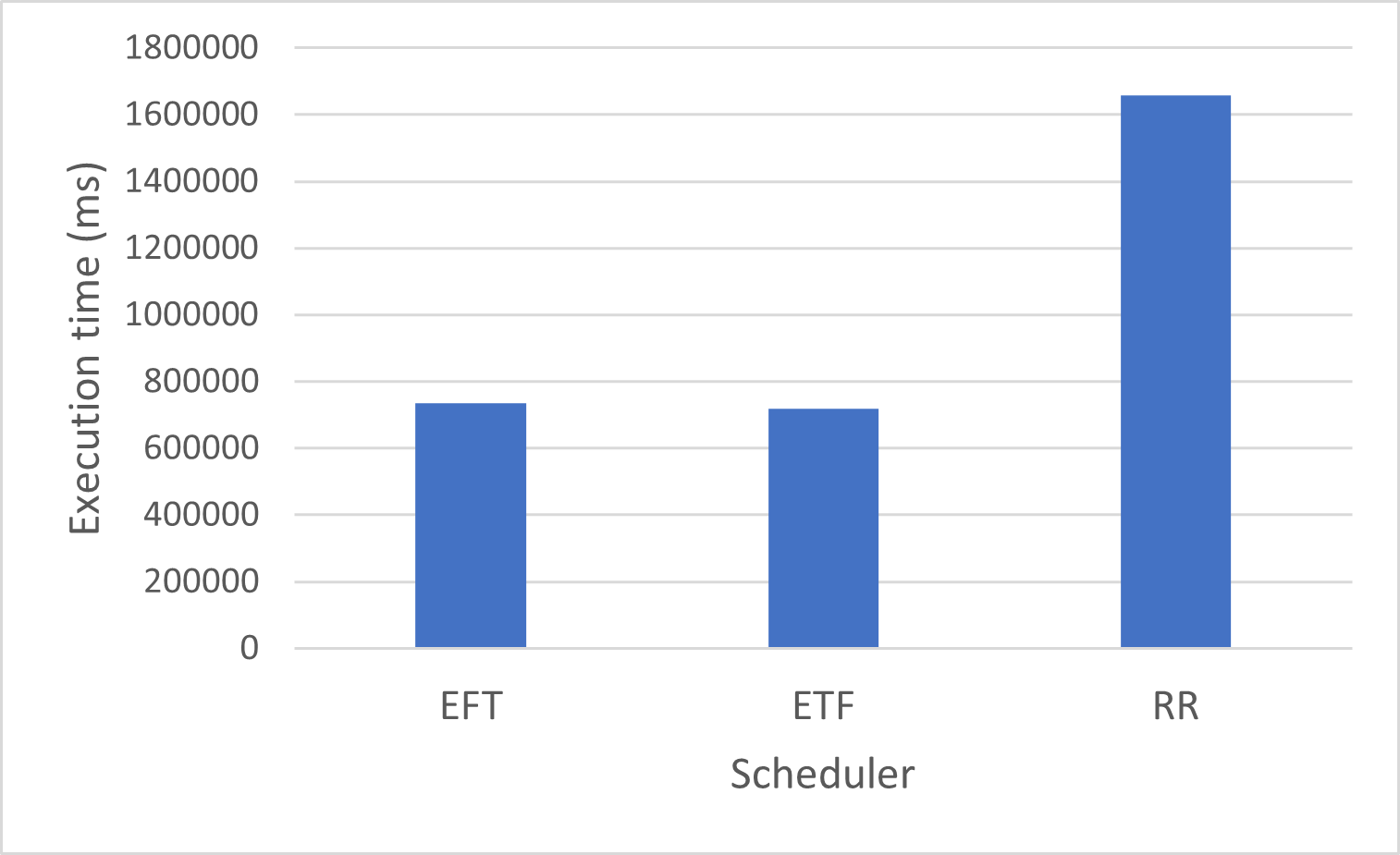}}    
    \hspace{1cm}
    \subfigure[]{\includegraphics[width=0.45\textwidth]{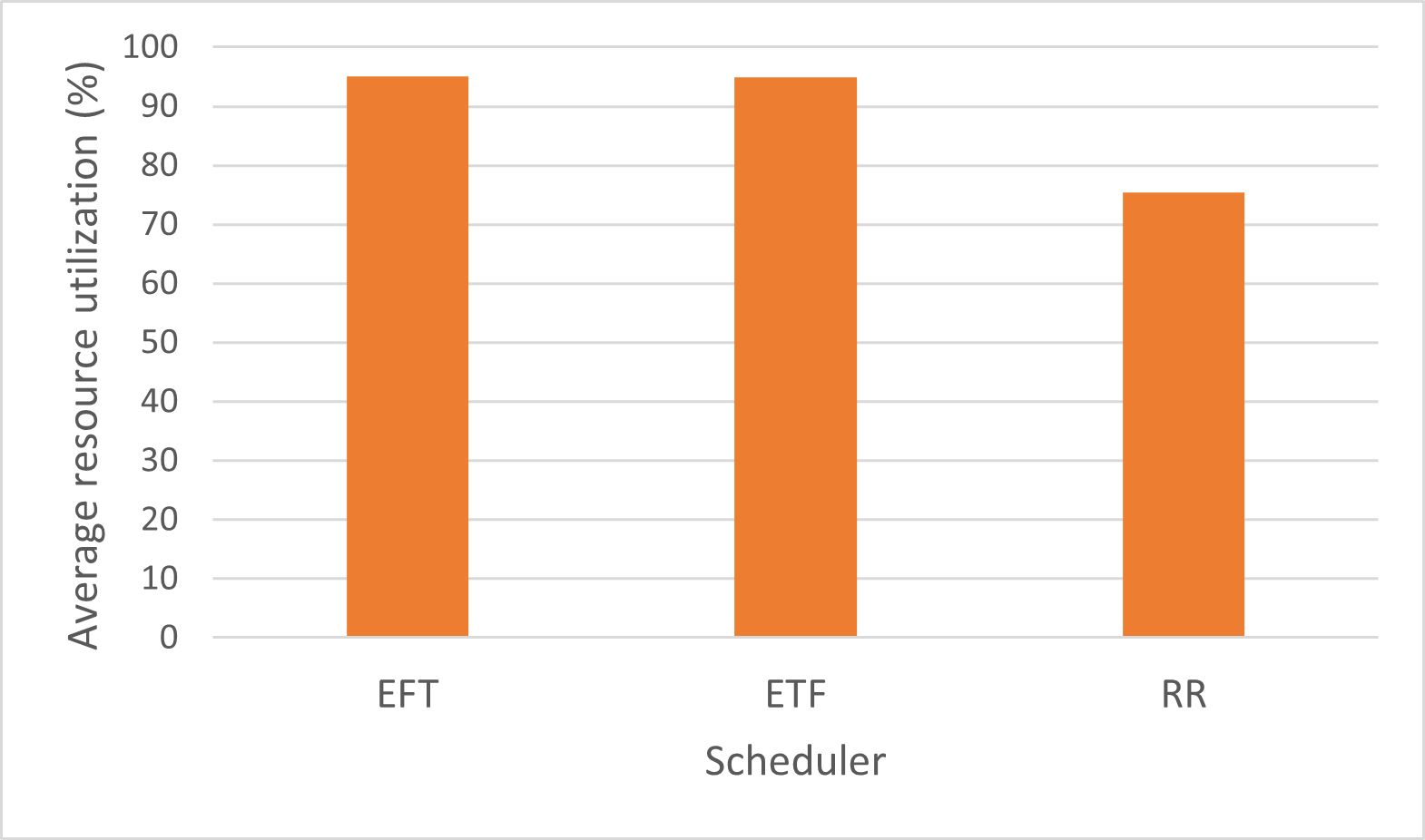}}
    \caption{Resource pool configuration: 3 ARM CPU cores, 1 Nvidia Volta GPU (Frontend), 3 Xeon CPU Cores, 1 Nvidia Tesla GPU and 1 Xilinx Alveo FPGA (Backend). Using EFT, ETF and RR schedulers, (a) presents execution time of 100 DS-workflow jobs, and (b) presents average resource utilization.}
    \label{fig:scheduler-sweep}
\end{figure}

Figure~\ref{fig:scheduler-sweep}(a) and (b) presents bar plots with the schedulers plotted along X-axis, and the execution time and mean resource utilization plotted along Y-axis respectively. Both these figures show that ETF and EFT schedulers perform very closely in terms of execution time and mean resource utilization. From Figure~\ref{fig:scheduler-sweep}(a), we notice that both ETF and EFT schedulers reduce the execution time by around 57\% compared to the RR scheduler. Furthermore, looking at Figure~\ref{fig:scheduler-sweep}(b), we notice that both ETF and EFT schedulers increase the mean resource utilization by upto around 21\% compared to the RR scheduler. These results demonstrate that the runtime using hierarchy aware schedulers can significantly improve the execution time and mean resource utilization compared to the simpler scheduling heuristics.

\subsubsection{Discussion}
In the JITA-4DS environment, the resource management problem can be  more complex and requires the design of new heuristics. The computing resources allocated to the VDC for a given class of applications are a heterogeneous mixture of different processing devices (different CPUs, different GPUs, different accelerators, etc.) with various execution performance and energy consumption characteristics. They depend on each of the specific applications being executed by that VDC. Our future work will consider this variables in the decision making process.
For example, several aspects remain open, like the ad-hoc design of the JITA-4DS resource management system for a  VDC built from a fixed set of components. The design of a JITA-4DS instance is determined by the execution time and energy consumption cost, and resources requirements of a data science pipeline. Therefore, it is necessary to dynamically identify the configuration choices for a given pipeline and define VDC resources' effective resource allocation strategies.   In general, for determining the dynamic resources requirements of data science pipelines at runtime, it is necessary to consider two challenges.
First, calculate a VDC-wide Value of Service (VoS) for a given interval of time,  weigh individual values of various instances of pipelines. We have started providing a study on VoS for JITA-4DS \cite{akoglu2021putting}, we will integrate these observations in further experiments on the current emulation.  
Second, propose objective functions that can guide heuristics to operate in the large search space of resource configurations. The objective is to derive possible adequate allocations of the shared and fixed VDC resources for several instances of data science pipelines.
 We have observed that decisions must be made regarding the resource management system for JITA-4DS to divide the shared, fixed resource pool across different VDCs to maximize the overall system-wide VoS. 
 All of the above single VDC challenges still apply and interact across VDCs. Additional problems, such as determining when resources should be reallocated across VDCs and do so in an online fashion, must be addressed. This includes the methodologies for reassigning resources that do not interfere with currently executing applications on different VDCs affected by the changes and measuring and accounting for the overhead of establishing new VDC configurations.

\section{Conclusion and Future Work}\label{sec:conclusion}
This paper introduced JITA-4DS, a virtualized architecture that provides a disaggregated data center solution ad-hoc for executing DS pipelines requiring elastic access to resources. DS pipelines process big streams and data coordinating operators implemented by services deployed on edge. Given that operators can implement greedy tasks with computing and storage requirements beyond those residing on edge, they interact with VDC services. We have set the first simulation setting to study resources delivery in JITA-4DS.

We are currently addressing challenges of VDCs management in simpler environments, on cloud resource management heuristics, 
big data analysis, 
and data mining for performance prediction. We give the first experimental results of these aspects.
To simulate, evaluate, analyze, and compare different heuristics, we will further build simulators for simpler environments and combine open-source simulators for different levels of the JITA-4DS hierarchy. We are currently defining a "benchmark" with different types of data science workloads that share data collections and functions and study complex resource allocation patterns.

%
%
%
 \bibliography{biblio}

 \end{document}